# Resonant Semiconductor Metasurfaces for Generating Complex Quantum States

**Authors:** Tomás Santiago-Cruz[1,2]*, Sylvain D. Gennaro[3,4], Oleg Mitrofanov[3,5], Sadhvikas Addamane[3,4], John Reno[3,4], Igal Brener[3,4], and Maria V. Chekhova[1,2]

**Affiliations:**

[1]Max Planck Institute for the Science of Light; Staudtstraße 2, 91058 Erlangen, Germany

[2]Friedrich-Alexander-Universität Erlangen-Nürnberg; Staudtstraße 7/B2, 91058 Erlangen, Germany.

[3]Center for Integrated Nanotechnologies, Sandia National Laboratories; Albuquerque, New Mexico 87185, USA.

[4]Sandia National Laboratories; Albuquerque, New Mexico 87185, USA.

[5]University College London; WC1E 7JE, London, UK.

*Corresponding author. Email: jose-tomas.santiago@mpl.mpg.de

**Abstract:** Quantum state engineering, the cornerstone of quantum photonic technologies, mainly relies on spontaneous parametric down-conversion and four-wave mixing, where one or two pump photons decay into a photon pair. Both these nonlinear effects require momentum conservation (i.e., phase-matching) for the participating photons, which strongly limits the versatility of the resulting quantum states. Nonlinear metasurfaces, due to their subwavelength thickness, relax this constraint and extend the boundaries of quantum state engineering. Here, we generate entangled photons via spontaneous parametric down-conversion in semiconductor metasurfaces with high-quality resonances. By enhancing the quantum vacuum field, our metasurfaces boost the emission of photon pairs within narrow resonance bands at multiple selected wavelengths. Due to the relaxed momentum conservation, the same resonances support photon pair generation from pump photons of practically any energy. This enables the generation of complex frequency-multiplexed quantum states, in particular cluster states. Our results demonstrate the multifunctional use of metasurfaces for quantum state engineering.

Quantum state engineering with photons mainly relies on nonlinear optical effects such as spontaneous parametric down-conversion (SPDC) or spontaneous four-wave mixing (SFWM). These effects create a vast variety of photonic quantum states, including single (*1*) and entangled (*2*) photons, squeezed (*3*) and cluster (*4–6*) states. However, both SPDC and SFWM in conventional nonlinear crystals and waveguides require strict momentum conservation for the involved photons, which strongly limits the versatility of the states they produce. The emergent concept of quantum optical metasurfaces helps to overcome this constraint. Metasurfaces, i.e., arrays of nanoscale resonators, feature unique abilities to manipulate and control the amplitude, phase, and polarization of light in the nonlinear (*7–9*) and quantum regimes (*10*) using a single ultrathin device. In particular, all-dielectric metasurfaces made of materials with high second-order nonlinearities have offered a potential route for on-chip quantum state generation (*11–13*) and manipulation. Due to the subwavelength thickness of metasurfaces, the momentum conservation (or phase-matching) requirement is relaxed (*14*), enabling multiple nonlinear processes to occur with comparable efficiencies (*15*). In addition, optical resonances in metasurfaces enhance the vacuum field fluctuations at certain wavelengths, boosting the spontaneous emission of photons (*16*). The same effect enhances the spontaneous emission of photon pairs via SPDC in optical antennas (*17*) and metasurfaces (*11*, *13*).

Vacuum field enhancement scales with the quality (Q) factor of the resonance. In metasurfaces, Q-factors are especially high for bound states in the continuum (BIC) resonances (*18–20*), which are discrete-energy modes whose energy levels overlap with a continuous spectrum of radiating modes (*21*). In symmetry-protected BIC metasurfaces, the outcoupling of radiation in the normal direction is forbidden by symmetry (*22*). As a consequence, the Q factors of these modes are infinite, and could in theory infinitely enhance the spontaneous emission of photons and photon pairs (*23*). In practice, the enhancement is finite due to symmetry breaking (quasi-BICs) (*22*) but can be still as high as $10^2$-$10^4$.

In this work, we report on the experimental generation of tunable photon pairs via spontaneous parametric down-conversion driven by high-Q BIC resonances in gallium arsenide (GaAs) quantum optical metasurfaces (QOMs). We show how our QOMs emit frequency-degenerate and non-degenerate narrowband photon pairs that can be tuned over more than a hundred nm by changing either the optical pump or the spectral location of the BIC without appreciable loss of efficiency. Moreover, by the appropriate choice of optical resonances and pump wavelengths, we can drive as many as necessary multiple SPDC processes simultaneously, leading to the emission of frequency multiplexed entangled photons. Our work paves the way for building room-temperature nanoscale sources of complex tunable entangled states for quantum networks.

In SPDC, a pump photon of a higher frequency $\omega_p$ down-converts in a second-order nonlinear material into a pair of signal and idler photons of lower frequencies, $\omega_s$ and $\omega_i$, following energy conservation (Fig. 1A). Contrary to thick nonlinear crystals, SPDC in subwavelength sources does not require longitudinal momentum conservation leading to the broadband emission of photon pairs over a wide range of angles (24, 25). In optical nanoantennae and metasurfaces, however, the resonances select the range of wavelengths and wavevectors where the photon emission is enhanced (11, 13, 17). Therefore, under judicious choice and design of optical modes and resonances, metasurfaces can be used to generate tunable, unidirectional photons at the nanoscale.

To demonstrate experimentally the SPDC enhancement with BICs, we fabricated various arrays of broken-symmetry resonators arranged in a square lattice of different periodicities and scaling, via standard electron beam lithography and chlorine-based dry etching. Subsequent fabrication steps include epoxy bonding and curing, substrate lapping and polishing, wet etching, and transferring the metasurface onto a transparent fused silica substrate. (See Supplementary Table 1 for various metasurface dimensions.) We choose GaAs as it possesses one of the highest second-order susceptibilities among traditional materials, between $\chi^{(2)} = 400 - 500$ pm.V$^{-1}$, for the range of pump wavelengths involved in this work; these susceptibilities exceed the nonlinear coefficients of ferroelectric nonlinear materials such as lithium niobate by more than one order of magnitude (26). The structure of the metasurfaces, as well as the crystallographic axes orientation, is shown in Fig. 1B.

The existence of symmetry-protected BICs can be explained through symmetry breaking and coupling between allowed and forbidden optical Mie modes (27), or using group theory arguments (28). A metasurface consisting of square meta-atoms obeys C$_2$ and C$_4$ rotational symmetry; photons at the BIC frequency are trapped inside the resonators and cannot escape into free space due to zero coupling to radiating modes. A small notch in the cube breaks the rotational symmetry (Fig. 1C) and transforms these symmetry-protected BICs into quasi-BICs, which can outcouple to the far field. Spectrally, they appear as narrow transmission peaks in the white-light far-field transmittance (Fig. 1D). The modes, labelled electric dipole (ED)-BIC and magnetic dipole (MD)-BICs, reach quality factors of $Q_{BIC\,(ED)} \approx$ 330 and $Q_{BIC\,(MD)} \approx 1000$, respectively. The MD-BIC and ED-BIC have different coupling efficiencies with respect to the incident beam polarization as shown in Supplementary Figure S1. For the lowest-order quasi-BICs, their simulated electric field profiles resemble those of out-of-plane dipole modes (see insets in Fig. 1D). Additional field distribution cross sections can be found in Supplementary Figure S2.

By tuning the period of the array and the proportions of the resonators, the central wavelengths of the BICs can be tuned over a wide range of wavelengths (Supplementary Figure S3). For emission normal to the QOM, the Q factor is the highest since most of the system symmetry is preserved. However, off-normal emission breaks the symmetry and the Q factor of each BIC decreases, lowering the field enhancement. Thus, we should also expect the emitted SPDC photon pairs to be radiated unidirectionally along the metasurface normal as observed in the past for emission from quantum dots embedded inside a symmetry-protected quasi BIC resonator (*16*).

To demonstrate multiplexed entangled photons generation via SPDC, we investigate three QOMs, labelled, respectively QOM-A, QOM-B and QOM-C, with different resonator sizes and resonance spectral spacings, such that the ED-BIC and MD-BIC optical modes are resonant at different wavelengths (Fig. 2A-C, upper panels). All metasurfaces are pumped with a linearly polarized continuous-wave tunable laser focused down to a spot size of 140 μm ($1/e^2$). Photon pairs are registered as joint detection events using two superconducting nanowire single-photon detectors placed at the two outputs of a fiber beam splitter (see Supplementary Figure S4). For all QOMs considered, we register a high number of simultaneous photon detections – coincidences (Fig. 1E) - which indicates the presence of photon pairs. Further details of the experimental apparatus are given in the Supplementary Information. Supplementary Figure S5 also shows the results of the second-harmonic generation, which is mediated by the same nonlinear tensor as SPDC and is therefore helpful for investigating the SPDC polarization properties.

We then perform two-photon fiber-assisted spectroscopy with 3 nm resolution to find the spectrum of the emitted photons. By placing a 3 km single-mode dispersive fiber into the path of the photons, we map the time delay between these photons to their wavelengths. The three corresponding SPDC spectra are shown in Fig. 2A-C (bottom panels), each one below its white-light transmission spectrum. Thanks to the relaxed momentum conservation, the pump wavelength $\lambda_p$ can be arbitrary, and thus we use different pumps and different QOMs to demonstrate various types of photon pairs that can be generated in our system.

For QOM-A, we use $\lambda_p = 723$ nm, such that the degenerate wavelength $2\lambda_p$ overlaps with the ED-BIC resonance wavelength. In this case, both SPDC photons are emitted within a single narrow peak, centered at the resonance wavelength ($2\lambda_p = 1446$ nm), see Fig. 2A. The full width at half maximum (FWHM) of 4.3 nm matches the linewidth of the ED-BIC resonance. This corroborates previous observations on QOMs (*11*) that the presence of an optical resonance enhances the quantum vacuum field within the resonance's bandwidth.

However, very different spectra are obtained when we pump QOM-B at 718 nm, such that the degenerate wavelength $2\lambda_p$ is off-resonance from the ED-BIC modes. Now, we find that the SPDC spectrum exhibits two narrow peaks, one centered at the ED-BIC peak wavelength ($\lambda_s = 1391$ nm), and the other at a wavelength $\lambda_i = 1485$ nm (Fig. 2B). This indicates the first-time observation of non-degenerate photon pairs emitted from QOMs. Here, the ED-BIC resonance enhances the vacuum field at the signal wavelength, forcing the QOM to emit a photon at this wavelength simultaneously with its partner at the idler wavelength, as dictated by energy conservation: $\lambda_i = \left(\frac{1}{\lambda_p} - \frac{1}{\lambda_s}\right)^{-1}$.

The spectrum of the entangled photons gets richer when both ED-BIC and MD-BIC resonances are active in SPDC (Fig. 2C). We achieve this by pumping our third metasurface QOM-C at 725 nm, with the pump radiation polarized at 40 deg, so that the MD-BIC resonance at 1429 nm is activated (see the Supplementary Information). Now, we observe four peaks, two of them corresponding to signal photons emitted at the ED and MD resonances and the other two, to their idler partners. We find that, since the MD-BIC optical resonance has a higher Q-factor than the ED-BIC resonance, its emission FWHM is narrower, causing the signal and idler SPDC spectra to also be narrower.

We also find that our QOMs can produce degenerate and non-degenerate photon pairs over a broad spectral range without considerable reduction of efficiency (Fig. 2D), and, due to the high-Q resonances, with an efficiency of at least three orders of magnitude higher than in an unpatterned GaAs film of the same thickness (see the Supplementary Information).

To verify the non-classicality of our photon pairs, we measure the pump-power dependence of the second-order cross-correlation function at zero time delay, defined as $g_{si}^{(2)}(0) = \frac{\langle \hat{N}_s \hat{N}_i \rangle}{\langle \hat{N}_s \rangle \langle \hat{N}_i \rangle}$, where $\hat{N}_{s,i}$ are photon-number operators for the signal and idler modes. The cross-correlation function is measured as $g_{si}^{(2)}(0) = \frac{R_c}{R_s R_i T_c}$, where $R_{c,s,i}$ are the rates of the signal-idler coincidences, signal photon detections, and idler photon detections, respectively, and $T_c$ the coincidence resolution time (*29*). For SPDC, all three numbers $R_{c,s,i}$ scale linearly with the pump power, which leads to the inverse dependence of $g_{si}^{(2)}(0)$ with power, as indeed observed in our measurement for QOM-A (Fig. 3A). While a strong correlation peak in $g_{si}^{(2)}(0)$ (Fig. 1D) is generally accepted as the evidence of nonclassical behavior, a formal proof requires the violation of the Cauchy-Schwarz (CS) inequality. Specifically, for classical fields the CS inequality holds (*30*),

$$\left(g_{si}^{(2)}(0)\right)^2 \leq g_{ss}^{(2)}(0) \cdot g_{ii}^{(2)}(0),$$

(1)

where $g_{ss,ii}^{(2)}(0)$ is the second-order autocorrelation function for the signal and idler modes, respectively, defined as $g_{ss}^{(2)}(0) = \frac{\langle :\hat{N}_s^2: \rangle}{\langle \hat{N}_s \rangle^2}$, with :: denoting normal ordering. Note that the CS inequality (1) can be tested only when the modes *s* and *i* are distinguishable. Since QOM-B provides two distinguishable frequency modes, at 1391 nm and 1485 nm, we measure the autocorrelation functions of modes *s* and *i* after 50 nm FWHM bandpass filters centered at 1400 nm (s) and 1475 nm (i), respectively. Figures 3B,C show the measured autocorrelation functions for the *s* and *i* modes, respectively. The single count graphs for each measurement are shown in the Supplementary Information. We get $g_{ss}^{(2)}(0) = 1.6 \pm 0.3$, and $g_{ss}^{(2)}(0) = 1.2 \pm 0.2$. The cross-correlation function is measured without bandpass filters (Fig. 3D), which gives $g_{si}^{(2)}(0) = 10.5 \pm 1.1$. Hence, the CS inequality is violated by >50 standard deviations, $\left(g_{si}^{(2)}(0)\right)^2 = 110 \pm 2, g_{ss}^{(2)}(0) \cdot g_{ii}^{(2)}(0) = 1.9 \pm 0.3$, revealing the nonclassical character of the photons generated in our QOMs.

We emphasize that the presence of narrow optical resonances enables the creation of entangled states and more complicated graph quantum states from our metasurfaces, as illustrated in figure 4A. In the current setup (see Supplementary Figure S4), the state preparation relies on the fiber beam splitter splitting the pairs between the two detectors; events where both photons arrive at the same detector are ignored. For a non-degenerate photon pair at wavelengths $\lambda_{1,2}$, after the beam splitter photon $|\lambda_1\rangle$ is path-entangled with photon $|\lambda_2\rangle$. If a second non-degenerate pair at wavelengths $\lambda_{1,3}$, arrives at the beam splitter, photon $|\lambda_1\rangle$ is also path-entangled with photon $|\lambda_3\rangle$ after it. Note that this photon pair needs another pump for its generation. If photons $|\lambda_1\rangle$ from each pair are indistinguishable, a linear graph state of three pairwise entangled qubits appears (Fig. 4A, top). This strategy of cluster state generation, called pairwise coupling (*6*), can be implemented using two coherent pump beams, at wavelengths 725 nm and 718 nm, and a resonance at 1359 nm (Fig. 4B). In our experiment, the two pump beams are incoherent, but they still provide the desired spectrum.

By adding a third pump, generating photon pairs at wavelengths $\lambda_{1,4}$, a more complicated graph state can be created, called a Greenberger-Horne-Zeilinger state (*5*), shown in Fig. 4A (middle).

The state becomes increasingly complex by adding multiple coherent pump beams at different wavelengths - for instance, using a frequency comb or a filtered supercontinuum as an excitation source. By appropriately matching the wavelength separation of the comb, and the optical resonances of the QOMs, photons at multiple wavelengths can be entangled via pairwise couplings. With this approach, one could implement a scalable cluster state, needed for one-way quantum computation (*5*). An example of such a state is shown in Fig. 4A (bottom). We stress that such methods of quantum state engineering are made possible by the use of our unique QOMs with the relaxed phase matching and engineered high-Q resonances, and are impossible to create using bulk crystal or waveguide SPDC sources. Moreover, QOMs provide a unique way to spatially multiplex multiple metasurfaces within the area of a single pump beam, by exciting resonances at different wavelengths (Fig. 4C), and entangling multiple photon pairs across separate wavelengths with a single pump.

In summary, we have demonstrated, for the first time, SPDC driven by quasi-BIC optical resonances in QOMs. The metasurface design allows for the generation of photon pairs at multiple wavelengths, with either one or both photons of a pair created at the wavelength of the BIC resonance. Unlike in previously studied Mie-type QOMs, there is no considerable loss or efficiency with passing from degenerate to non-degenerate SPDC. The resulting state features entanglement; in the case of pumping coherently at different wavelengths, scalable graph quantum states can be generated. In contrast to other on-chip sources of photon pairs, the QOM geometry offers additional degrees of freedom, such as transverse displacement of various structures within a single pump beam, opening up vast opportunities for on-chip quantum state engineering.

**Acknowledgments:**

T.S.C. and M.V.C. are part of the Max Planck School of Photonics, supported by BMBF, Max Planck Society, and Fraunhofer Society.

T.S.C. and M.V.C. acknowledge support by the Deutsche Forschungsgemeinschaft (DFG, German Research Foundation) – Project ID 429529648 – TRR 306 QuCoLiMa ("Quantum Cooperativity of Light and Matter"). S.D.G., O.M. and I.B. were supported by the U.S. Department of Energy, Office of Basic Energy Sciences, Division of Materials Sciences and Engineering. The work was performed, in part, at the Center for Integrated Nanotechnologies, an Office of Science User Facility operated for the U.S. Department of Energy (DOE) Office of Science. Sandia National Laboratories is a multi-mission laboratory managed and operated by National Technology and Engineering Solutions of Sandia, LLC, a wholly owned subsidiary of Honeywell International, Inc., for the U.S. Department of Energy's National Nuclear Security Administration under contract DE-NA0003525. This paper describes objective technical results and analysis. Any subjective views or opinions that might be expressed in the paper do not necessarily represent the views of the U.S. Department of Energy or the United States Government.

We thank A. V. Rasputnyi, V. Sultanov, and M. Poloczek for their help at the initial stage of the experiment, and M. Lippl and N. Y. Joly for giving us access to the tunable continuous-wave laser.

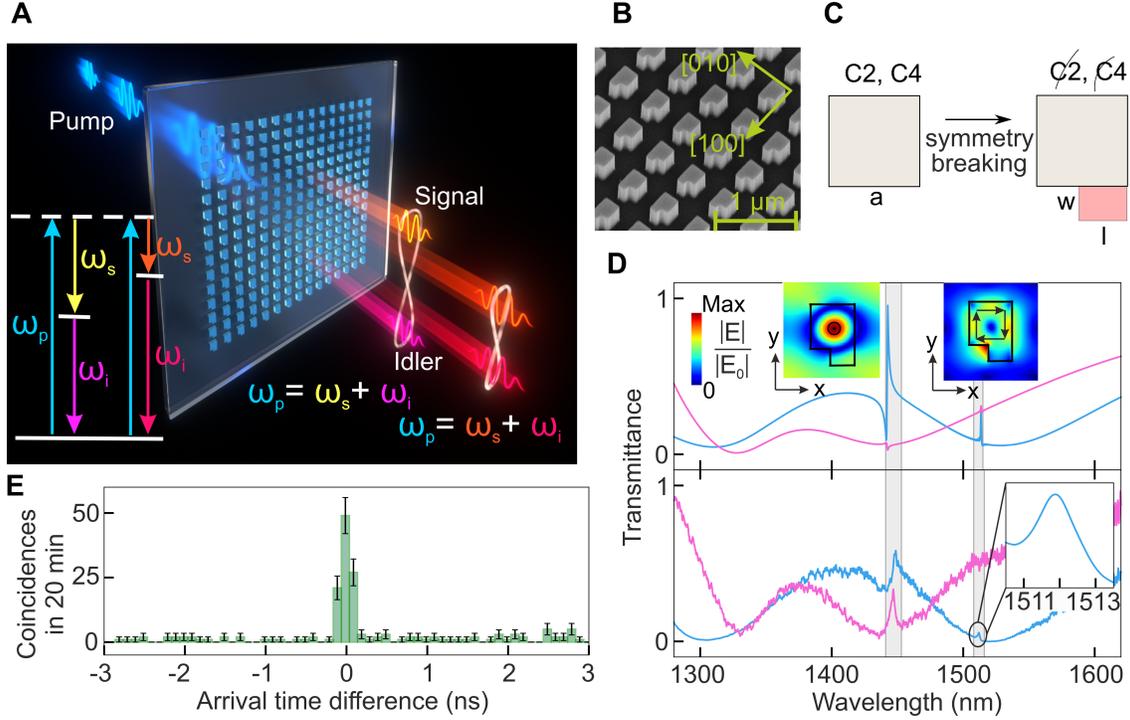

**Fig. 1. SPDC using symmetry-protected quasi-BIC resonances in a semiconductor metasurface**. (**A**) Conceptual diagram of multiplexed entangled photon generation in a multi-resonance semiconductor metasurface. (**B**) Scanning electron micrograph of the metasurface at an intermediate step of nanofabrication (see SI). (**C**) Structure of the broken-symmetry resonators. The parameters are: height 500nm, large square width $a = 320$ nm, small rectangle width $w = 115$ nm, length $l = 173$ nm. The base periods of the array vary from 0.47 $\mu$m to 0.55 $\mu$m, which tunes and redshifts the BIC resonances. The addition of a small rectangle (pink) breaks the symmetry of the metasurface, turning the BIC into a quasi-BIC, which can outcouple to the far field. C2 and C4 denote rotational symmetries. (**D**) Simulated (top) and measured (bottom) white-light linear transmission spectra of one metasurface considered in this work, for two incident polarizations: along the [110] direction (blue) and orthogonal to it (pink). Grey areas highlight the locations of the quasi – BIC resonances: the ED-BIC ($\lambda = 1446$ nm) and the MD-BIC ($\lambda = 1512$ nm). The inset shows the top view (xy) of the electric-field normalized distribution calculated at the center of the nanoresonator for both quasi-BIC resonances. Dark arrows indicate the direction of the electric field. (**E**) Typical distribution of the time difference between the photon arrivals at two detectors, demonstrating photon pair generation.

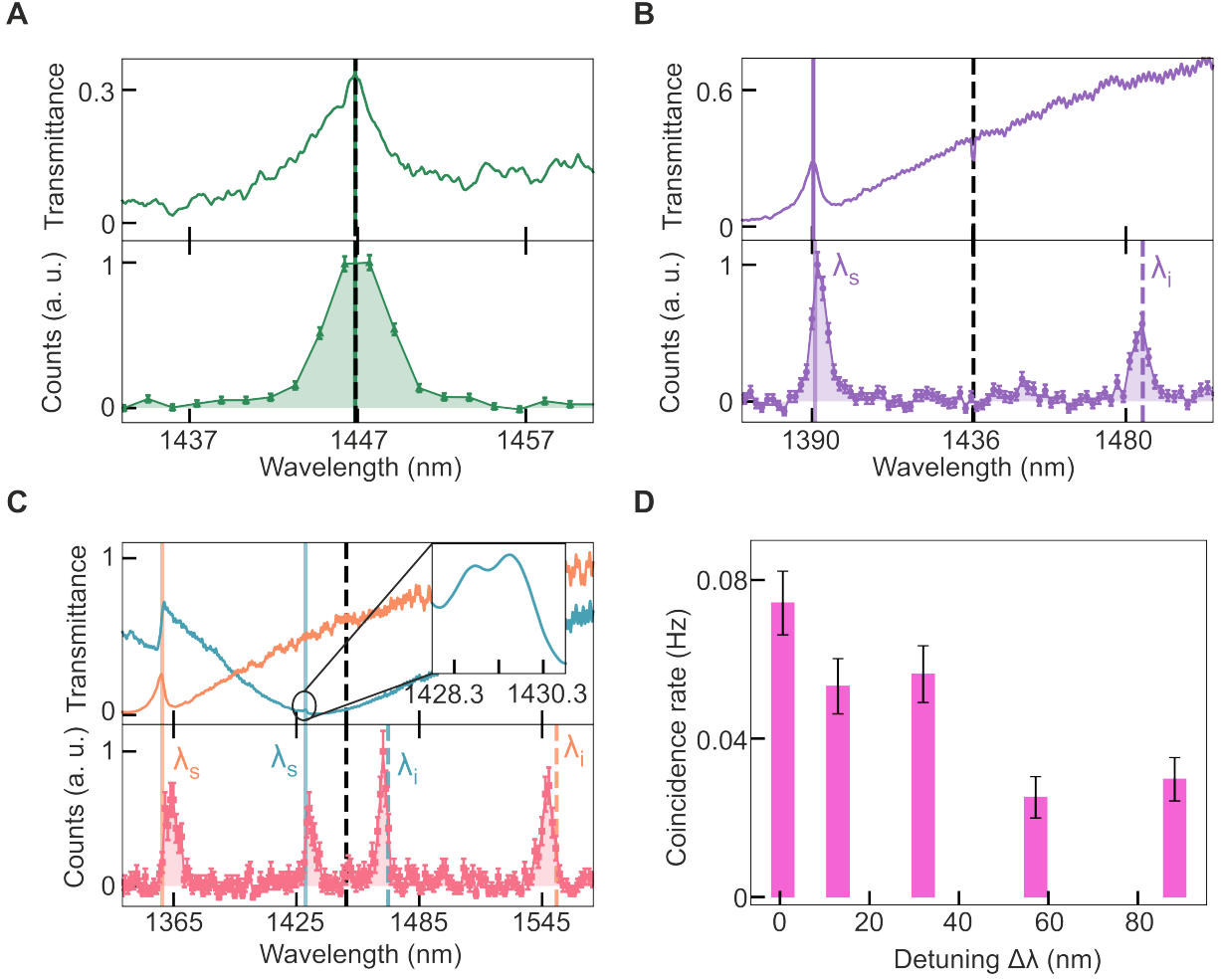

**Fig. 2. SPDC spectra for the QOMs considered in this work. (A – C)** Measured white-light transmission spectra (top panels) and SPDC spectra measured via the fiber-assisted spectroscopy (bottom panels) for QOMs A – C, respectively. The SPDC spectra show (A) the production of degenerate photon pairs at the wavelength $2\lambda_p = 1446$ nm (dark dashed line), which overlaps with the electric dipole (ED)-BIC resonance of QOM-A; (B) the production of non-degenerate photon pairs where only the signal photon is emitted at the ED-BIC mode wavelength $\lambda = 1391$ nm (purple vertical solid line) of QOM-B; (C) the production of two types of non-degenerate photon pairs, where signal photons are emitted at wavelengths $\lambda = 1359$ nm of the ED-BIC resonance (orange vertical line) and $\lambda = 1429$ nm of the magnetic dipole (MD)-BIC resonance (green vertical solid line) of QOM-C. Both ED- and MD-BIC modes are active in SPDC due to the choice of the pump polarization. **(D)** Coincidence rate as a function of the wavelength detuning $\Delta\lambda = 2\lambda_p - \lambda_{ED-BIC}$ between the degenerate wavelength and the ED-BIC resonance, $\lambda_{ED-BIC}$ of five QOMs. The pump power in this experiment was 9.6 mW.

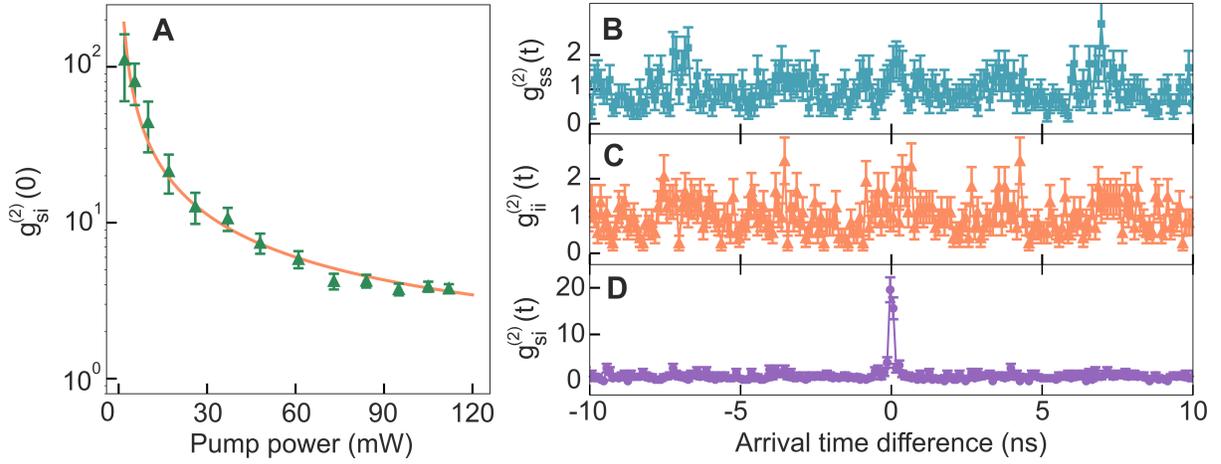

**Fig. 3. Second-order cross- and auto-correlation functions.** (**A**) Pump power dependence of the second-order auto-correlation function at zero-time delay measured for photon pairs emitted by QOM-A. The solid line is a fit of the expected dependence. (**B,C**) Second-order auto-correlation functions $g^{(2)}_{ss}$ and $g^{(2)}_{ii}$ as measured for QOM-B at the signal wavelength (1391 nm) and idler wavelength (1485 nm), respectively. (**D**) Second-order cross-correlation function $g^{(2)}_{si}$ as measured for QOM-B.

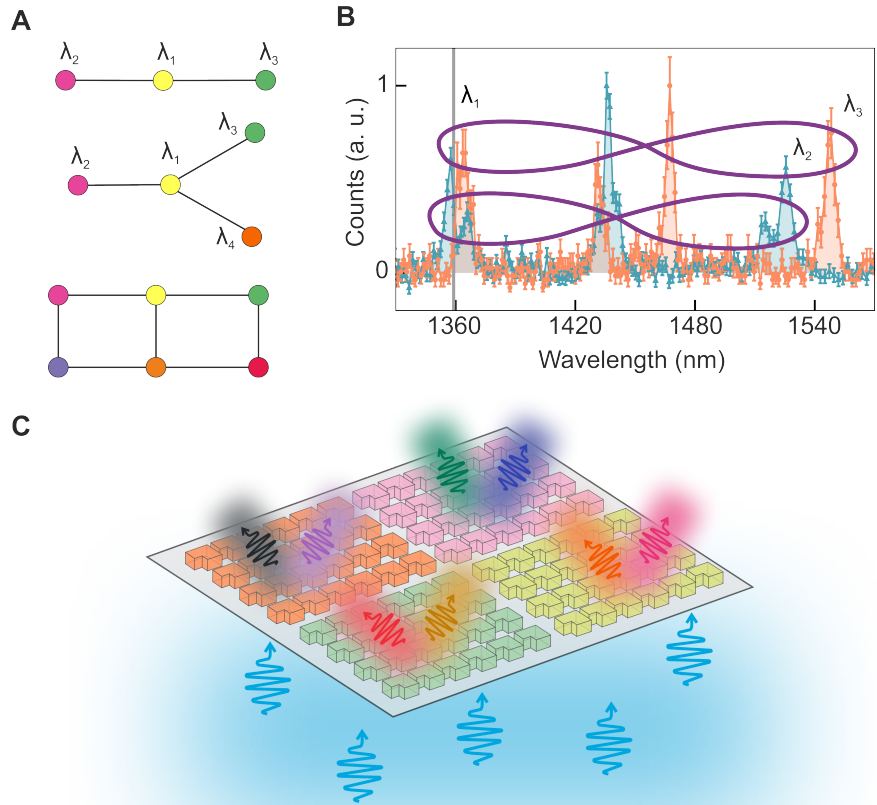

**Fig. 4. Cluster state generation with QOMs.** (**A**) Three examples of cluster states: a linear three-qubit graph state (top), a Greenberger-Horne-Zeilinger state (middle) and a more general graph state (bottom). (**B**) SPDC spectrum illustrating how a linear three-qubit graph state of photons $|\lambda_1\rangle, |\lambda_2\rangle, |\lambda_3\rangle$ can be generated from QOM-C. (**C**) Spatial multiplexing of various (4 shown) metasurfaces for cluster state generation using a single pump beam.